\documentclass[12pt]{iopart}
\usepackage{iopams,bm}
\usepackage{cite}

\usepackage{color,soul}
\usepackage{graphicx}

\newcommand{\reallylongrightarrow}{-\!\!\!-\!\!\!\longrightarrow}

\definecolor{orange}{RGB}{255,127,0}
\begin{document}

\title[Stochastic dynamics on slow manifolds]{Stochastic dynamics on slow manifolds}
\author{George W A Constable$^1$,  Alan J McKane$^1$, Tim Rogers$^2$}
\address{$^1$ Theoretical Physics, School of Physics and Astronomy, The University of Manchester, Manchester M13 9PL, UK}
\address{$^2$ Centre for Networks and Collective Behaviour, Department of Mathematical Sciences, University of Bath, Bath, BA2 7AY, UK}
\begin{abstract}
The theory of slow manifolds is an important tool in the study of deterministic dynamical systems, giving a practical method by which to reduce the number of relevant degrees of freedom in a model, thereby often resulting in a considerable simplification. In this article we demonstrate how the same basic methodology may also be applied to stochastic dynamical systems, by examining the behaviour of trajectories conditioned on the event that they do not depart the slow manifold. We apply the method to two models: one from ecology and one from epidemiology, achieving a reduction in model dimension and illustrating the high quality of the analytical approximations. 
\end{abstract}
\maketitle

\section{Introduction}

\sethlcolor{orange}

It is possibly only a slight exaggeration to say that of all the mathematical models we can dream of, there are only two kinds which are straightforward to solve: those which are linear, and those which are one-dimensional. This aphorism holds equally for stochastic dynamical systems as it does for their deterministic counterparts. Much of applied mathematics and theoretical physics is devoted to the delicate art of taking high-dimensional or non-linear problems of interest and finding appropriate approximation schema with which to reduce their apparent difficultly. 

The theory of `slow' or `centre' manifolds is an example of just such a scheme and one which is well developed for deterministic dynamical systems \cite{Wiggins}. In many models of interest there exists a separation of time scales between some quantities which relax very quickly to an essentially static value, and others which change more slowly and can be sensitive to perturbations. The term `slow manifold' describes the space in which these slower quantities vary, after any fast initial transient has died out. Restricting attention to this space offers an effective method by which to remove the so-called fast degrees of freedom, thereby reducing the dimension and simplifying the system. In this article we propose a new method in which the principles of slow manifold theory from deterministic systems can be used in stochastic systems derived from an individual based model (IBM) as a tool to separate time scales and reduce the dimensionality of the model. Importantly we find that the nature of the  mesoscopic stochastic system is strongly constrained by the individual level dynamics of the IBM in such a way that the analysis may be simplified.

The goal of removing fast degrees of freedom from stochastic systems has received significant attention and a variety of approximation methods have previously been proposed \cite{stratanovichHaken,itoHaken,elimGardiner,hutt1,hutt2,KnoblochWiesenfeld,Titulaer,Grima,serra,Contou-Carrere,Wang,Lan,RogersMcKaneRossberg2012,coullet,arnoldNormalForm,arnoldNormalFormCenterMan,sriNamaNormalForm,roberts,pavBook}, though a simple and generally applicable theory is yet to emerge. The existing literature on the subject can be coarsely split according to the framework within which the stochastic system is represented. Some authors have focused on master or Fokker-Planck type equations which govern the time evolution of the probability distribution of system states \cite{Gardiner}. Such a view was perhaps first pioneered by Knobloch and Wiesenfeld \cite{KnoblochWiesenfeld}. Within this formalism, a reduction of dimension can be achieved through the application of projection operators, as illustrated by Gardiner \cite{elimGardiner}, and developed by others \cite{Titulaer,Grima}. Further work moving away from such projection methods has also been conducted \cite{hutt1,hutt2} utilising the rigorous work by Boxler on stochastic slow manifolds\cite{boxler}. Unfortunately, the details of these methods can be somewhat cumbersome, especially in cases where the fast degrees of freedom are not parallel to the variables with which the system is described. The alternative formalism of stochastic differential equations (SDEs) provides a perhaps more intuitive approach. The equations describe the evolution of trajectories themselves and bear a useful resemblance to ordinary differential equations, which aids physical reasoning. For this reason we choose the SDE formulation as the basis for our work. However, it should be pointed out that such systems cannot be correctly interpreted without specifying the choice of stochastic calculus, and that certain manipulations are not entirely straightforward \cite{ito}. 

Perhaps the simplest implementation of time scale separation in the SDE setting is the treatment of `direct adiabatic elimination' presented in \cite{serra}. It will be instructive to review the salient points. The procedure mirrors closely that of slow manifold theory: the variables associated with the fast dynamics are assumed to be stationary, from which a function describing the slow manifold may be determined. {If the fast variables are subject to noise, then the procedure will yield an expression for the slow manifold which contains noise variables; unfortunately this can lead to ill-defined terms \cite{serra}. The method can, however, be usefully applied in systems where the entire system is linear in the fast variables. The more rigorously derived and well-known Haken slaving principle \cite{wunderlinHaken,stratanovichHaken,itoHaken}, and several other related methods \cite{Contou-Carrere,Wang,Lan} are developed along similar lines of reasoning and suffer from the same complications which arise from specifying the slow manifold in a stochastic sense. A notable exception is \cite{Doering}, which deals with a particular model in which there is a true centre manifold (that is, a surface on which there is no deterministic flow), and applies a novel method in which stochastic perturbations away from the manifold are assumed to instantaneously relax along the deterministic trajectories. 

More mathematically rigorous work on SDE fast variable elimination has been conducted for stochastic analogues of normal form coordinate transformations. While perhaps the earliest example in the SDE setting was \cite{coullet}, work has been significantly extended in the intervening years \cite{arnoldNormalForm,arnoldNormalFormCenterMan,sriNamaNormalForm,arnoldBook}. However, many of these transformations result in noise convolutions which involve anticipating future unknown noise terms. Further work has proposed the use of additive noise terms to emulate these convolution terms in the limit of long times \cite{khasminskii,sriNamaEquivAvNormForm,chaoRobertsLowDimModel}, though these methods are arguably less formal than the theory they rest within. Perhaps the most significant advance in this area therefore came in \cite{roberts}, with the construction of a methodology for a stochastic normal form transform that avoids such anticipating memory integrals in many cases, though even here there remain many situations where a long time additive noise substitution must be invoked. A body of work also exists on averaging and homogenisation techniques \cite{pavBook,berglund}, although both have a more limited range of applicability than stochastic normal forms \cite{roberts} and the former has been shown in certain cases to be equivalent to a stochastic normal form \cite{sriNamaEquivAvNormForm}. One of the biggest drawbacks of the work on normal forms however is that it almost exclusively deals in SDE systems with uncorrelated noise terms, whereas SDEs derived from an underlying microscopic models often exhibit strong noise correlations.

Individual based models have recently become very popular in several areas of physics and applied mathematics as tools to study complex emergent phenomena, where they are used to examine the effect of demographic noise. In the limit of large system size, the mesoscopic behaviour of a given IBM is well described by an SDE system derived from the rules of the model (see \cite{McKaneBiancalaniRogers2013} and the appendices of this paper for details). In addition to often having correlated noise terms they also have a range of other features that make them distinct from generic SDEs. It is to these types of models that we will apply our method, exploiting some of their unique features.

The core of our approach is to examine the behaviour of a stochastic system in the SDE framework under the condition that its trajectories are \textit{confined} to the slow manifold of the deterministic version of the system. We have applied a similar procedure in several previous works \cite{RogersMcKaneRossberg2012,RogersMcKaneRossberg2012a,McKaneBiancalaniRogers2013}, with successful results. We note that our aim is not the recreation of individual stochastic trajectories, but rather the preservation of the system's statistical properties. Because we use a static description of the slow manifold, the method is applicable to a broader range of systems derived from IBMs than the direct elimination procedure or the Haken slaving principle. Moreover, the procedure is mathematically explicit, straightforward to apply, and addresses the effect of correlated noise terms. One also gains a sense of physical intuition as to the behaviour of the system, which is arguably not present in the master equation setting.

We will also show how the slow manifold approximation can be effectively combined with other stochastic approximation techniques. The linear noise approximation (LNA, or van Kampen expansion \cite{vanKampen}) has found favour amongst theoreticians studying IBMs. In a nutshell, the LNA provides a mesoscopic description of the system in terms of a linear SDE. The price paid for this simplification is that the theory only applies in the neighbourhood of an attractive fixed point of the noise-free version of the model. In the context of the LNA, slow manifolds can be a malign presence. If some eigenvalues of the approximate linear SDE are close to zero, then a small stochastic fluctuation in the direction of the corresponding eigenvector can carry the system very far from the steady state into regions in which the true non-linear nature of the model is important. Moreover, a large separation between eigenvalues can in some situations lead to the numerical evaluation of theoretical solutions becoming ill-conditioned. Both of these effects can lead to a very poor agreement between stochastic simulations and the LNA theory. 

In the next section we develop our method with the aid of a simple illustrative example with an ecological interpretation, before providing a general formulation. The results from this example demonstrate the success of the approximation scheme even in regimes where the fixed point is weakly unstable; this addresses the first difficulty with slow manifolds and the LNA identified above. In section three we go on to apply the general formulation of the method to an epidemiological model with seasonal forcing. This model has been identified as suffering from the technical numerical difficulties associated with a large separation between eigenvalues \cite{Ganna}. We show how our method may be used in tandem with the LNA to provide a very good approximation to results coming from stochastic simulations.

\section{Method}
\subsection{Motivation}
We begin by recapping the basics of slow manifolds in deterministic systems. Consider the ordinary differential equation 
\begin{equation}
\frac{d\bm{x}}{dt} = \bm{A}(\bm{x})\,,
\label{det}
\end{equation}
where $\bm{x}$ is an $n$-dimensional vector describing the state of the system, and $\bm{A}$ is an $n$-dimensional vector-valued function of $\bm{x}$. As is well known, the behaviour of the system in the neighbourhood of a fixed point $\bm{x}^\ast$ is described by the linearization of $\bm{A}$ around that point. Define the Jacobian matrix $J$ with entries
\begin{equation}
J_{ij}=\frac{\partial}{\partial x_j}A_i(\bm{x})\bigg|_{\bm{x}=\bm{x}^\ast}\,.
\end{equation}
Then for $\bm{x}$ close to $\bm{x}^\ast$, the time evolution of $\bm{\xi}=\bm{x}-\bm{x}^\ast$ obeys 
\begin{equation}
\frac{d\bm{\xi}}{dt} = J\bm{\xi}\,.
\label{det_err}
\end{equation}
Further insight is gained by considering the eigenvalues and eigenvectors of $J$. From (\ref{det_err}), we learn that if $\bm{v}$ is a right eigenvector of $J$ with eigenvalue $\lambda$, then errors in the direction of $\bm{v}$ will grow exponentially if $\textrm{Re}[\lambda]>0$ and shrink exponentially if $\textrm{Re}[\lambda]<0$. If, on the other hand, $\textrm{Re}[\lambda]=0$ then we do not know what effect perturbations in the direction of $\bm{v}$ will have on the long-term behaviour of the system. To answer this question would require a more detailed non-linear analysis, which is likely to be very difficult in a general system with several degrees of freedom. Slow manifold theory offers a way to make progress in this case by reducing the dimension of the model. 

The basic observation behind the theory is as follows: since perturbations in the direction of stable/unstable eigenvectors will shrink/grow exponentially, the only trajectories whose behaviour is in question are those which are tangential to the span of the eigenvectors with eigenvalue zero. Very often, this set of eigenvectors has many fewer members than there are degrees of freedom in the original system, and thus restricting attention to this subspace achieves a considerable reduction in dimensionality. 

Slow manifolds are also of great practical use when no eigenvalues are precisely zero, but there is a separation of time-scales. For example, suppose a stable fixed point $\bm{x}^\ast$ has associated eigenvalues satisfying $\textrm{Re}[\lambda_1]<\ldots<\textrm{Re}[\lambda_{m}]\ll\textrm{Re}[\lambda_{m+1}]<\ldots<0$. Perturbations in the direction of eigenvectors $\bm{v}_{1}\,,\ldots\,,\bm{v}_{m}$ will decay extremely rapidly in comparison with those in the directions of $\bm{v}_{m+1}\,,\ldots\,,\bm{v}_n$. For practical purposes, the `slow' manifold of trajectories tangent to these less stable eigenvectors defines an $(n-m)$-dimensional system which will provide a good qualitative approximation to the behaviour of the larger system, as perturbations away from this manifold will very quickly collapse.

Our goal in this article is to apply the basic ideas of slow manifolds to stochastic systems. Our starting point will be the SDE
\begin{equation}
\frac{d\bm{x}}{dt} = \bm{A}(\bm{x})+\bm{\eta}(t)\,,
\label{sde}
\end{equation}
where $\bm{x}$ and $\bm{A}$ are as in (\ref{det}), and $\bm{\eta}$ is a vector of Gaussian white-noise variables with zero mean and correlations
\begin{equation}
\Big\langle\eta_i(t)\eta_j(t')\Big\rangle=\varepsilon\,\delta(t-t')B_{ij}(\bm{x})\,.
\label{corr}
\end{equation}
Here $\langle\cdots\rangle$ denotes averaging over the noise, $\varepsilon$ is a small parameter governing the strength of the noise, $B$ is a matrix-valued function of the system state, and the SDE (\ref{sde}) is to be interpreted in the It\={o} sense. We also draw attention to our choice of notation: the vector $\bm{\eta}(t)$ here contains correlated noise terms, as opposed to being a vector of uncorrelated Gaussian variables as used by some authors. The exact form of $\bm{A}$ and $B$ will be determined by the application.

We are interested in the case where the deterministic ($\varepsilon=0$) system exhibits a slow manifold. How will the stochastic system behave if we confine its trajectories to this slow manifold? We develop the theory with the aid of a specific example.

\subsection{Illustrative example}\label{toyModel}
To illustrate our method, we explore the behaviour of a simple ecological model of two interacting populations, labelled $X$ and $Y$. Individuals of both populations reproduce with rate one, and there is a small probability $\mu$ of the offspring mutating from one type to the other. The organisms also prey on each other with rate $\varepsilon$ and a slight preference $p$ for prey of the opposite type. The model may be written in the traditional notation of chemical reaction systems:

\begin{eqnarray}\label{ecomodel}
\textrm{Reproduction:}\qquad\quad &X\stackrel{1-\mu}{\reallylongrightarrow}X+X\,,\quad Y\stackrel{1-\mu}{\reallylongrightarrow}Y+Y\nonumber\\ \nonumber\\ 
\textrm{Mutation:} &X\stackrel{\mu}{\reallylongrightarrow}X+Y\,,\quad Y\stackrel{\mu}{\reallylongrightarrow}X+Y\nonumber\\ \nonumber\\ 
\textrm{Predation:} & X+Y\stackrel{\varepsilon (1/2+p)}{\reallylongrightarrow}X\,,\quad X+Y\stackrel{\varepsilon (1/2+p)}{\reallylongrightarrow}Y\nonumber\\\nonumber \\
\textrm{Cannibalism:} & X+X\stackrel{\varepsilon (1/2-p)}{\reallylongrightarrow}X\,,\quad Y+Y\stackrel{\varepsilon(1/2-p)}{\reallylongrightarrow}Y\,.\\\nonumber 
\end{eqnarray}
Here arrows denote possible reactions and the values above them are the rate constants. Writing $n_X$ and $n_Y$ for the number of individuals in each population, the model may be mathematically formulated as a master equation describing the time evolution of the probability distribution $P(n_X,n_Y)$; see \eref{Meqn} in Appendix A. Stochastic simulations of the model can be performed efficiently using the Gillespie algorithm \cite{gillespie}. 

When the predation rate $\varepsilon$ is small, the population may grow very large. Performing an expansion of the master equation in the limit of small $\varepsilon$ yields an effective description of the system in terms of an SDE for the scaled variables $x=\varepsilon n_X$ and $y=\varepsilon n_Y$. We give the details in Appendix A. For the present model, we find the following pair of equations:
\begin{eqnarray}
\label{xysde}
\frac{dx}{dt}=x-\mu(x-y) - x\left(\frac{1}{2}(x+y)-p(x-y)\right)+\eta_x(t)\,,\nonumber\\ \\
\frac{dy}{dt}=y+\mu(x-y) - y\left(\frac{1}{2}(x+y)+p(x-y)\right)+\eta_y(t)\,,\nonumber
\end{eqnarray}
where $\eta_x$ and $\eta_y$ have the correlation structure specified in (\ref{corr}), with 
\begin{equation}
\label{xyB}
\fl B=\left(\begin{array}{cc}x+\frac{1}{2}x(x+y)-(px+\mu)(x-y)&0\\0&y+\frac{1}{2}y(x+y)+(py+\mu)(x-y)\end{array}\right)\,.
\end{equation}

We begin by examining the deterministic system found by putting $\varepsilon=0$. There is a trivial fixed point at $x^{*}=0\,,\,y^{*}=0$, representing the extinct state, which is always unstable. There is a second fixed point at $x^{*}=1\,,\,y^{*}=1$, representing equal coexistence of the two populations. This state is stable when $p<\mu$. If $p$ is raised above $\mu$, a supercritical pitchfork bifurcation occurs, with the equal coexistence fixed point becoming unstable and giving rise to a symmetric pair of stable fixed points in which one species dominates the other. The new fixed points have coordinates
\begin{equation}
\fl x^{*}=\frac{1-2\mu\pm\sqrt{(1-\mu/p)(1-2\mu)}}{1-2p}\,,\,\,y^{*}=\frac{1-2\mu\mp\sqrt{(1-\mu/p)(1-2\mu)}}{1-2p}\,.
\end{equation}
We are interested in examining the effect of noise near this transition. 

The eigenvalues of the Jacobian at the coexistence state are $-1$ and $\lambda \equiv 2(p-\mu)$, with corresponding eigenvectors $(1,1)$ and $(1,-1)$. If $|\lambda|\ll1$ then we have a slow manifold in the direction of $(x-y)$, meaning that perturbations to the balance of populations evolve very slowly. Formally, the slow manifold is defined by the collection of trajectories which are tangent to the slow eigenvector at the fixed point, although in practice there is unlikely to be a closed analytic expression for this surface. To make progress we approximate the slow manifold by the surface on which the rate of change in the direction of the fast eigenvector is zero (known as the nullcline). In the present model, the nullcline of the fast direction $(x+y)$ is given by $dx/dt + dy/dt = 0$ which yields the hyperbola
\begin{equation}
(x+y)-\frac{1}{2}(x+y)^2+p(x-y)^2=0\,.
\label{nclxy}
\end{equation}
\begin{figure}
\begin{center}
\includegraphics[width=0.4\textwidth]{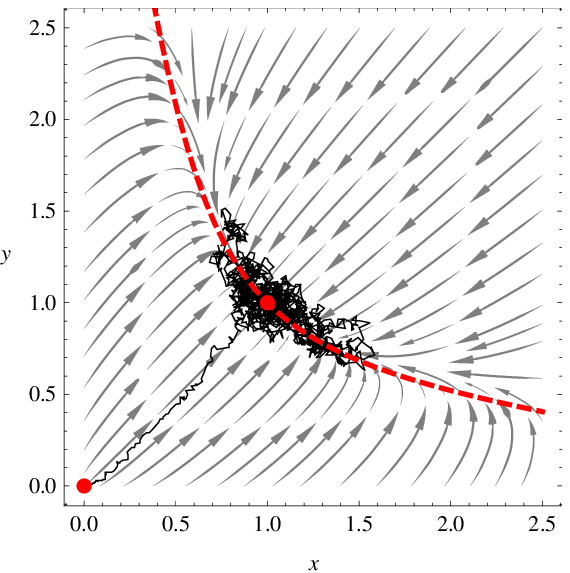}\hspace{15mm}
\includegraphics[width=0.4\textwidth]{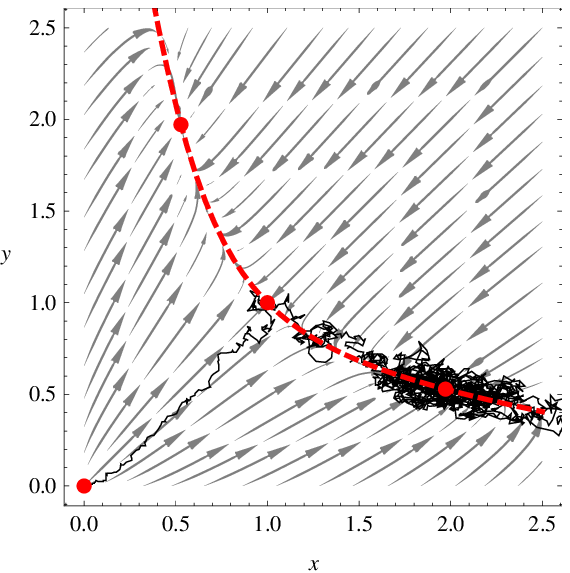}
\end{center}
\caption{These plots show the behaviour of our illustrative ecological model on either side of the pitchfork bifurcation. The fixed points are shown as red circles, and the dashed red line is the nullcline $dx/dt+dy/dt=0$, given in  \eref{nclxy}. The grey arrows show trajectories of the deterministic system, while the black line traces out the trajectory of a single (short) stochastic simulation of the individual-based model, starting close to the origin. The parameters are $\varepsilon=0.005$ and $p=0.3$ in both plots, while $\mu=0.35$ on the left and $\mu=0.25$ on the right.  }\label{toy_fig_1}
\end{figure}
The two plots in \fref{toy_fig_1} capture the typical behaviour of the model for parameters either side of the transition. 

The SDE system (\ref{xysde}) is two-dimensional, non-linear and has noise correlations which depend on the state of the system. These factors combine to make the theoretical analysis of the model very difficult. The situation is not hopeless, however, as it is clearly visible in \fref{toy_fig_1} that the system state does not typically stray very far from the one-dimensional subspace defined by the nullcline of the fast variable $(x+y)$. We intend to exploit this fact to produce an `effective' one-dimensional description of the model. The plan of attack is as follows: first we will make a coordinate transform to separate the fast and slow variables; then we will examine the behaviour of the slow variable under the assumption that the fast variable relaxes instantaneously to its nullcline value.

We introduce $w=x+y$ and $z=x-y$, so that 
\begin{equation}
\left(\begin{array}{c}w\\z\end{array}\right)=\left(\begin{array}{cc}1&1\\1&-1\end{array}\right)\left(\begin{array}{c}x\\y\end{array}\right)\, \equiv V \left(\begin{array}{c}x\\y\end{array}\right).
\label{transform}
\end{equation}
In the new coordinates the nullcline is described by the equation $w-w^2/2+pz^2=0$, and \eref{xysde} becomes
\begin{eqnarray}
\label{wzsde}
\frac{dw}{dt}=w-\frac{1}{2}w^2 +pz^2 +\eta_w(t)\,,\nonumber\\ \\
\frac{dz}{dt}=z\left(1-2\mu-\left(\frac{1}{2}-p\right)w\right)+\eta_z(t)\,.\nonumber
\end{eqnarray}
To determine the correlation structure of the new noise variables $\eta_w$ and $\eta_z$, we apply a general result on Gaussian random variables. Suppose that a vector of Gaussian random variables $\bm{\eta}$ has correlation matrix $B$, and that $\bm{\zeta}=V\bm{\eta}$ for some matrix $V$. What is the correlation matrix of $\bm{\zeta}$? Well,
\begin{equation}
\langle\zeta_i\zeta_j\rangle=\Big\langle\sum_{k,l}V_{ik}\eta_kV_{jl} \eta_l\Big\rangle=\sum_{k,l}V_{ik} \langle\eta_k\eta_l\rangle V^{T}_{lj}=\tilde{B}_{ij}\,,
\end{equation}
where $\tilde{B}=VBV^T$. In the present case, the matrices $B$ and $V$ are given in equations (\ref{xyB}) and (\ref{transform}), respectively. We thus find the following correlation matrix for $\eta_w$ and $\eta_z$:
\begin{equation}
\tilde{B}=\left(\begin{array}{cc}w+\frac{1}{2}w^2-pz^2&z(1-2\mu+w(\frac{1}{2}-p))\\z(1-2\mu+w(\frac{1}{2}-p))&w+\frac{1}{2}w^2-pz^2\end{array}\right)\,.
\label{corrwz}
\end{equation}
Notice that whilst the original noise variables $\eta_x$ and $\eta_y$ were independent (the off-diagonal entries of $B$ were zero), the noise variables in the new coordinates are correlated with each other.

To enforce the assumed separation of time-scales between $w$ and $z$, we impose the following conditions:
\begin{equation}
w=1+\sqrt{1+2pz^2}\,\quad\textrm{and}\quad\,\eta_w(t)\equiv0\,.\label{illustrativeTheta}
\end{equation}
The first of these sets $w$ to its value on the nullcline for a given $z$, whilst the second removes the possibility of any noise-induced fluctuations.  What effect do these constraints have on the evolution of $z$? First, we may substitute $w=1+\sqrt{1+2pz^2}$ into (\ref{wzsde}) to remove the dependence on $w$, thus
\begin{equation}
\frac{dz}{dt}=f(z)+\eta_z(t)\,,
\label{zsde}
\end{equation}
where
\begin{equation}
f(z)=z\left(1-2\mu-\left(\frac{1}{2}-p\right)\Big(1+\sqrt{1+2pz^2}\Big)\right)\,.
\end{equation}
Second, we must determine the effect of the conditions on the noise variable $\eta_z$. Since $\eta_w$ and $\eta_z$ are correlated, imposing $\eta_w=0$ will alter the statistical distribution of $\eta_z$. 

Again we apply a general result on correlated Gaussian random variables (see Appendix B of \cite{RogersMcKaneRossberg2012a}). Suppose that a collection of Gaussian random variables $(\eta_1,\ldots,\eta_n)$ has correlation matrix $\tilde{B}$. Let $\bar{B}$ be the correlation matrix of $(\eta_2,\ldots,\eta_n)$ conditioned on the event that $\eta_1=0$. Then $\bar{B}$ and $\tilde{B}$ are related by
\begin{equation}
\big[\bar{B}^{-1}\big]_{ij}=\big[\tilde{B}^{-1}\big]_{ij}\,,\quad\textrm{for all }\,\,i,j=2,\ldots, n\,.
\label{gauss_cond}
\end{equation}
This fact straightforward to prove by integration of the Gaussian probability density function. In particular, if $n=2$ then the variance of $\eta_2$ conditioned on $\eta_1=0$ is
\begin{equation}
\bar{B}=\tilde{B}_{22}-\frac{\tilde{B}_{12} \tilde{B}_{21}}{\tilde{B}_{11}}\,.
\label{gauss_cond2}
\end{equation}
Applying formula (\ref{gauss_cond2}) to the correlation matrix found in (\ref{corrwz}), we obtain 
\begin{equation}
\big\langle\eta_z(t)\eta_z(t')\big\rangle=\varepsilon\,\delta(t-t')g(z) \,,
\label{corrz}
\end{equation}
where the noise strength $g$ is given by 
\begin{equation}
g(z)=\left(w+\frac{1}{2}w^2-pz^2\right)\left(1 -\left(z \frac{1-2\mu+w(\frac{1}{2}-p)}{w+\frac{1}{2}w^2-pz^2}\right)^2\right)\,,
\end{equation}
and where of course $w=1+\sqrt{1+2pz^2}$. 

Equations (\ref{zsde}) and (\ref{corrz}) together define a one-dimensional stochastic differential equation. Although it may look a little complicated, being one-dimensional means that most questions of interest about this system can be answered by standard methods \cite{Gardiner}. For example, the long-time average behaviour of the model is captured in the stationary distribution of $z$, which has the following explicit form:
\begin{equation}
P(z)=\frac{1}{g(z)}\exp\left(\frac{2}{\varepsilon}\int_{-\infty}^z \frac{f(z')}{g(z')}\,dz'\right)\,.
\label{Pz}
\end{equation}
\begin{figure}
\begin{center}
\includegraphics[width=0.4\textwidth]{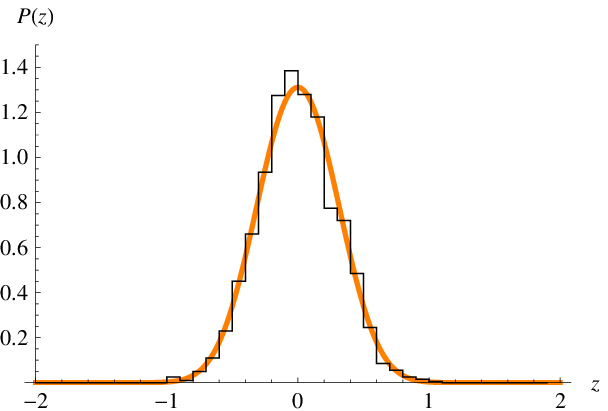}\hspace{15mm}
\includegraphics[width=0.4\textwidth]{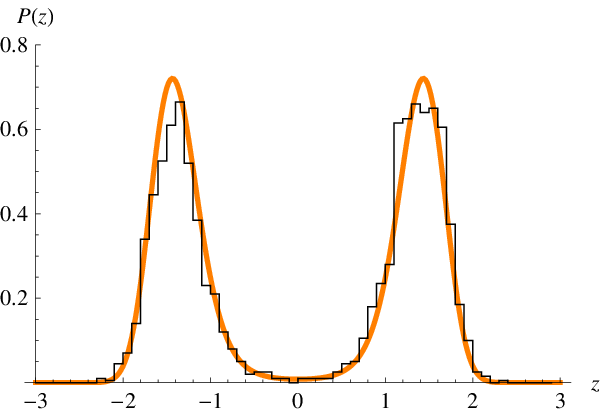}
\end{center}
\caption{The stationary distribution of $z=x-y$ as predicted by the reduced-dimension model (orange curve) and measured from a single long simulation run of the individual-based model (black histogram). The parameters are the same as those in \fref{toy_fig_1}. The theoretical prediction was obtained by numerically integrating \eref{Pz}, with parameters taken from \fref{toy_fig_1}, whilst 2000 data points were collected from simulations run at time intervals of 100.}\label{toy_fig_2}
\end{figure}
In \fref{toy_fig_2} we compare the analytical prediction of \eref{Pz} with a histogram of the $z$-coordinate of the sample points of stochastic simulations taken from \fref{toy_fig_1}. Clearly, the reduced one-dimensional model provides a very good fit to the data coming from the individual-based simulation. It should also be pointed out that although we have developed the theory based on the local behaviour around the coexistence fixed point $(1,1)$, the approximation remains successful even in the unstable regime.

\subsection{General formulation}\label{generalFormulation}
We close this section by providing a description of the method for an arbitrary $m$-dimensional IBM, whose dynamics are fully described by a set of $\ell$ reaction rates,
\begin{eqnarray}\label{generalReaction}
 \sum_{i=1}^{m} a_{ji}X_{i} \stackrel{r_{j}}{\reallylongrightarrow} \sum_{i=1}^{m}b_{ji}X_{i}, \quad \forall j=1, \dots \ell,
\end{eqnarray}
where $a_{ji}$ and $b_{ji}$ respectively specify the reactants and products of the $j^{th}$ reaction. The number of individuals in each population $X_{i}$ can then be expressed as $n_i$ in the manner described in \Sref{toyModel}. The model is then formulated as an $m$-dimensional master equation which now describes the evolution of the distribution $P(n_1 \dots n_{m})$; see \eref{Meqn} in \ref{stochasticBackground}. In most cases the total number of particles $N=\sum_{i}^{m}n_{i}$ is large; it is thus natural to expand in the small parameter $\varepsilon = 1/N$. Putting $x_{i}=\varepsilon n_{i}$ and expanding to second order in $\varepsilon$ one obtains the SDE
\begin{equation}
\frac{d\bm{x}}{dt} = \bm{A}(\bm{x})+\bm{\eta}(t)\,,
\end{equation}
with noise correlations
\begin{equation}\label{generalNoise}
\Big\langle\eta_i(t)\eta_j(t')\Big\rangle=\varepsilon\,\delta(t-t')\Big[B(\bm{x})\Big]_{ij}\,,
\end{equation}
where $i,j=1, \dots m$.

Suppose we are interested in behaviour around a fixed point $\bm{x}^\ast$. Let $J$ be the Jacobian of $\bm{A}$ at that point and write $\lambda_1\,,\ldots\,,\lambda_m$ for its eigenvalues. Suppose further that $\lambda_1$ is non-degenerate, real, large and negative, thus its associated eigenvector $\bm{v}_1$ represents a very stable direction. We aim to eliminate fluctuations in this direction to produce a reduced-dimension model. To apply our method, we first make a change of variables from the $m$-vector $\bm{x}$ to a single variable $w$ and an $(m-1)$-vector $\bm{z}$, via the coordinate transformation
\begin{equation}
{w \choose \bm{z}}=V\bm{x}\,.
\end{equation}
It is possible to choose $V$ so that 
\begin{equation}
\bar{J}=VJV^{-1}=\left(\begin{array}{cc}\lambda_1 & 0 \\0 &L\end{array}\right)\,,
\end{equation}
where $L$ is an $(m-1)\times(m-1)$ matrix. The first column of $V^{-1}$ must be $\bm{v}_{1}$, and therefore near the fixed point $\bm{x}^\ast$, the variable $w$ describes the distance from the slow manifold along the fast direction $\bm{v}_{1}$, while the remaining $m-1$ slow degrees of freedom are captured by $\bm{z}=(z_2,\ldots,z_m)^T$. In the new coordinates the SDE becomes
\begin{equation}
\frac{d}{dt}{w \choose \bm{z}} = V\bm{A}\left(V^{-1}{w \choose \bm{z}}\right)+\bm{\zeta}(t)\,,
\end{equation}
where
\begin{equation}
\Big\langle\zeta_i(t)\zeta_j(t')\Big\rangle=\varepsilon\,\delta(t-t')\Big[\tilde{B}(w,\bm{z})\Big]_{ij}\,,
\end{equation}
and $\tilde{B}=VBV^T$. We wish to constrain $w$ to its nullcline, which is defined by the equation
\begin{equation}
\bm{v}_{1}^T\bm{A}\left(V^{-1}{w \choose \bm{z}}\right)=0\,.
\label{nullc}
\end{equation}
To enforce the assumed separation of time-scales between $w$ and the other variables, we impose the following conditions:
\begin{equation}\label{ansatz}
w=\theta(\bm{z})\, \, \quad\textrm{and}\quad\,\zeta_1(t)\equiv0\,.
\end{equation}
In order to implement the method we need to obtain $\theta(\bm{z})$ analytically. However, beginning from an IBM of the type described in \eref{generalReaction} the resulting SDE system will be polynomial in the state variables, typically of low degree. When the solution for $\theta(\bm{z})$ is not unique one may frequently eliminate alternative solutions based on stability arguments. Notice that we are taking a static description of the slow manifold. While it has been shown that the deterministic slow manifold does not, in general, converge to the expectation of the stochastic slow manifold of generic SDEs \cite{roberts}, the discrepancy is typically of the same order as the noise. Since the noise in the IBM derived SDE model is small by construction \eref{generalNoise}, the deviation from the deterministic slow manifold will be negligible.

For the remaining variables, we have 
\begin{equation}
\frac{d\bm{z}}{dt} = \bm{\bar{A}}(\bm{z}) +\bm{\zeta}(t)\,,
\label{red_sde}
\end{equation}
where
\begin{equation}
\Big\langle\zeta_i(t)\zeta_j(t')\Big\rangle=\varepsilon\,\delta(t-t')\Big[\bar{B}(\bm{z})\Big]_{ij}\,, \quad \, \, i,j=2, \ldots m.
\label{red_cor}
\end{equation}
The drift vector $\bm{\bar{A}}(\bm{z})$ and diffusion matrix $\bar{B}(\bm{z})$ are derived from $\bm{A}$ and $\tilde{B}$ as follows. For $i,j=2,\ldots,m$ 
\begin{equation}
\Big[\bm{\bar{A}}(\bm{z})\Big]_{i}=\left[V\bm{A}\left(V^{-1}{\theta(\bm{z}) \choose \bm{z}}\right)\right]_{i}
\end{equation}
and 
\begin{equation}
\Big[\bar{B}(\bm{z})^{-1}\Big]_{ij}=\Big[\tilde{B}(\theta(\bm{z}),\bm{z})^{-1}\Big]_{ij}\,.
\label{Bpp}
\end{equation}
The general solution for the new noise covariance matrix can then be shown to be 
\begin{equation}
\bar{B}(\bm{z}) = \mathcal{B}_{22}(\bm{z}) - \mathcal{B}_{21}(\bm{z})  \mathcal{B}_{11}^{-1}(\bm{z})  \mathcal{B}_{12}(\bm{z}),
\end{equation}
where $\mathcal{B}$ are matrices of the partitioned $\tilde{B}$ matrix;
\begin{eqnarray}
\tilde{B}=  \left(\begin{array}{ccc} \mathcal{B}_{11}(\bm{z})  & \mathcal{B}_{12}(\bm{z}) \\  \mathcal{B}_{21}(\bm{z}) & \mathcal{B}_{22}(\bm{z}) \end{array}\right),
\end{eqnarray}
with $\mathcal{B}_{11}(\bm{z})$ a $1 \times 1$ matrix and $\mathcal{B}_{22}(\bm{z})$ an $(m-1) \times (m-1)$ matrix (see Appendix B of \cite{RogersMcKaneRossberg2012a}). Equations (\ref{red_sde}) and (\ref{red_cor}) describe a reduced-dimension stochastic system in which the fast direction associated with the eigenvalue $\lambda_1$ has been eliminated.

Finally, we note that it is possible to construct systems for which the conditioned dynamics are not qualitatively similar to those of the full system. In particular, if the noise matrix $B$ is singular, then eliminating the noise away from the manifold can lead to a system which is entirely deterministic. An example of such a system is discussed in \ref{singularExample}. This is not, however, a problem for SDEs arising from IBMs, as their noise matrices are generally non-singular, as discussed in \ref{stochasticBackground}. 

\section{Application: seasonally forced epidemics}\label{seirSection}
\subsection{Model definition and deterministic treatment}
The SEIR model is a simplified epidemiological model describing the spread of a disease through a population \cite{seirRef}. Members of the population may be in one of four states: susceptible ($S$), exposed ($E$), infectious ($I$) and recovered ($R$). The susceptible individuals come into contact with the infected and become exposed with infection rate $\beta(t)$, which may vary with time. Those exposed to the disease then become infectious with a rate of disease onset $\alpha$. Finally, the infectious recover with an average rate of $\gamma$. In addition to these disease dynamics, there is a constant birth and death rate $\mu$; it is traditional to hold the population size constant by treating death and birth as a single process whereby an individual returns to the susceptible state. As in the earlier illustrative model, the dynamics may be conveniently summarized using the notation of chemical reactions:
\begin{eqnarray}
\textrm{Infection:}\quad S + I\stackrel{\beta(t)}{\longrightarrow}E + I\qquad\qquad&\textrm{Incubation:} \quad E\stackrel{\alpha}{\longrightarrow}I\label{SEIRreactions}\\
\textrm{Recovery:}\quad I\stackrel{\gamma}{\longrightarrow}R&\textrm{Death/Birth:}\quad\, E,I,R\stackrel{\mu}{\longrightarrow}S\,.\nonumber
\end{eqnarray}
We write $n_S\,,n_E\,,n_I\,,n_R$ for the number of individuals in states $S$, $E$, $I$ and $R$, respectively. The total population size is then given by $N=n_S+n_E+n_I+n_R$, which does not vary, meaning that there are three degrees of freedom. With just a slight abuse of notation we introduce variables $S=n_S/N$, $E=n_E/N$ and $I=n_I/N$ which describe the population density of individuals in each disease state. Note that there is no need for a variable associated to the recovered state, since the conservation of total population makes it a dependent variable. Applying the same system-size expansion as before (see Appendix A for details), we obtain the following effective SDE system: 
\begin{eqnarray}
 \frac{dS}{dt}=\mu (1-S) -\beta(t) SI+\eta_1(t)\,,\nonumber\\
 \frac{dE}{dt}=\beta (t) SI - (\mu + \alpha)E+\eta_2(t)\,,\label{seirDet} \label{SEIsde}\\
 \frac{dI}{dt}=\alpha E -(\mu + \gamma)I+\eta_3(t)\,,\nonumber
\end{eqnarray}
where $\eta_{1,2,3}$ are Gaussian white noise variables with correlations
\begin{eqnarray}
&\Big\langle\eta_i(t)\eta_j(t')\Big\rangle=\frac{1}{N}\,\delta(t-t')B_{ij}\,,\nonumber\\
\nonumber\\
&B=\left(\begin{array}{ccc}\mu (1-S) + \beta(t)SI&-\mu E + \beta(t)SI&-\mu I\\-\mu E+ \beta(t)SI&\beta(t)SI + (\mu + \alpha)E&-\alpha E\\-\mu I&-\alpha E&\alpha E +(\mu + \gamma)I\end{array}\right)\,.
\label{SEIRcorr}
\end{eqnarray}

In this section we discuss the behaviour of the model in the deterministic limit $N\to\infty$, before moving on to consider the full stochastic system in \Sref{seirStochastic}. When the infection rate is not seasonally forced, so $\beta(t)\equiv\beta$, there are a pair of fixed points. The first of these represents the extinction of the disease: $S=1\,,E=0\,,I=0$. The second fixed point is
\begin{equation}
S^\ast=\frac{(\alpha+\mu)(\gamma+\mu)}{\alpha\beta}\,,\quad E^\ast=\frac{\mu(1-S^\ast)}{\alpha+\mu}\,,\quad I^\ast=\frac{\alpha\mu(1-S^\ast)}{(\alpha+\mu)(\gamma+\mu)}\,,
\end{equation}
and is referred to as the endemic state. We are concerned with the regime in which the endemic state is stable and the extinct state is unstable, which holds for a range of epidemiologically realistic parameter values. In general, the rate parameter $\mu$ controlling birth and death will be much smaller than the remaining rate parameters $\beta$, $\alpha$ and $\gamma$, since this takes place on a much longer timescale than the disease dynamics. The parameter $\mu$ can therefore be utilized as an expansion parameter to simplify some of the expressions in the analysis. To first order in $\mu$, the eigenvalues of the Jacobian at the endemic state are
\begin{equation}
\lambda_1=-(\alpha+\gamma)-\mu\,\frac{\alpha(2\alpha+\beta)+3\alpha\gamma+2\gamma^2}{(\alpha+\gamma)^2}\,,
\label{l1}
\end{equation}
and the complex-conjugate pair
\begin{equation}
\lambda_{2,3}= - \mu\,\frac{\alpha ^2 \beta+\beta \gamma ^2+\alpha  \gamma  (\beta +\gamma )}{2 \gamma  (\alpha +\gamma )^2} \pm i\, \sqrt{\mu\, \frac{\alpha  (\beta-\gamma )}{\alpha +\gamma }}\,.
\label{l23}
\end{equation}
Notice that we have a separation of time-scales: $\textrm{Re}[\lambda_1]\ll\textrm{Re}[\lambda_{2,3}]$, meaning that $\lambda_1$ corresponds to a highly stable direction. In addition, since $\mu$ is small, the imaginary parts of $\lambda_{2,3}$ are typically larger than the real parts, meaning that we may expect highly oscillatory trajectories in the neighbourhood of the endemic state. We can thus expect the system to first collapse rapidly in the direction of the first eigenvector, followed by a slow, almost-planar, spiralling decay to the endemic state, as shown in \fref{spirals}. This separation of timescales has been previously noted and exploited in the deterministic setting \cite{schwartz} and the stochastic setting using normal form techniques \cite{schwartzStoch}. However the stochastic analysis in \cite{schwartzStoch} considered a quite different system (additive noise on an unforced system which was not derived from a microscopic IBM), with very different objectives (namely the replication of stochastic trajectories).

\begin{figure}
\begin{center}
\includegraphics[width=0.4\textwidth]{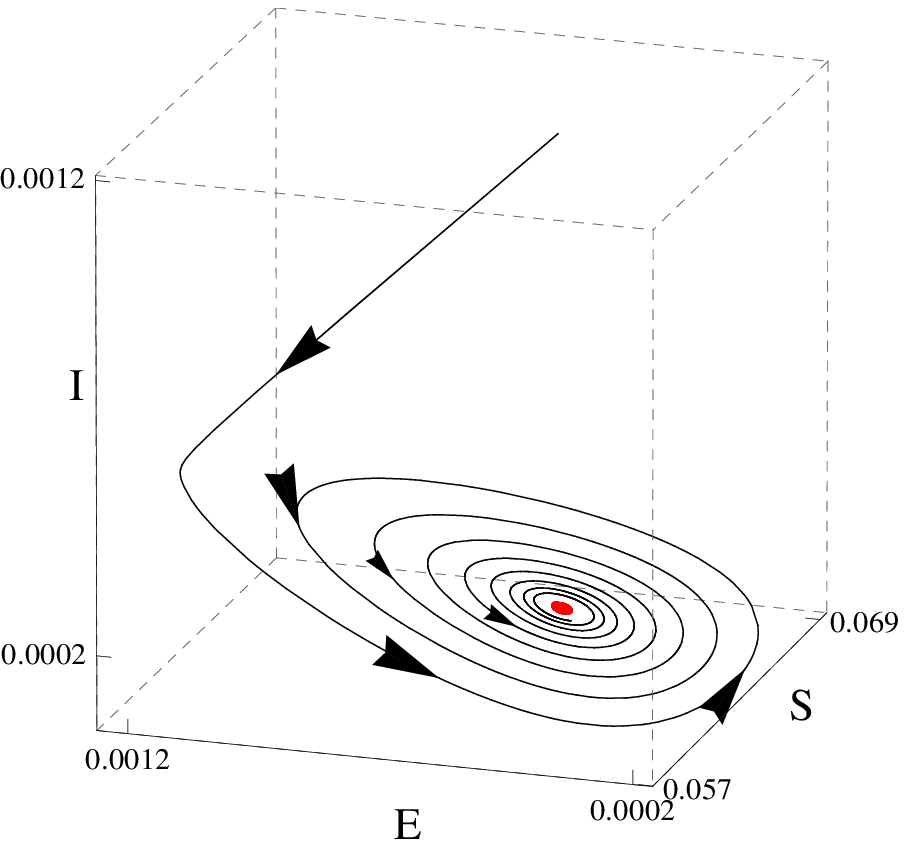}\hspace{15mm}
\includegraphics[width=0.4\textwidth]{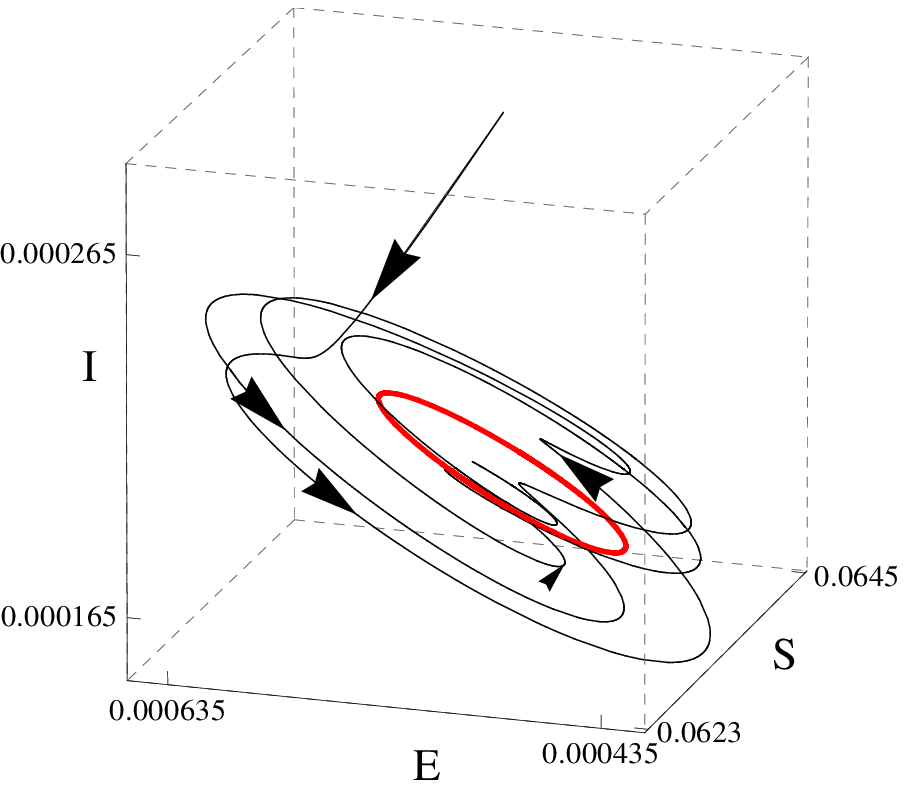}
\end{center}
\caption{Deterministic trajectories for the SEIR model. Left: No seasonal forcing ($\delta = 0$), fixed point shown in red. Right: system in the presence of forcing, ($\delta = 0.02$), limit cycle highlighted in red. Remaining parameters in both plots are $\beta_{0}=1575$ year$^{-1}$, $\alpha=35.84$ year$^{-1}$, $\gamma=100$ year$^{-1}$ and $\mu=0.02$ year$^{-1}$, which are standard choices for measles \cite{Ganna,SchwartzParameters}.}
\label{spirals}
\end{figure}

Introducing seasonal forcing of the infection rate creates an additional layer of complexity. A typical choice would be 
\begin{equation}
\beta(t)=\beta_0\Big(1+\delta\cos(2\pi t)\Big)\,,
\label{betat}
\end{equation}
where $\beta_0$ describes the basal infection rate, $\delta$ is the forcing amplitude, and time is measured in years. The deterministic system will now not settle to the endemic state, but instead exhibit limit cycle behaviour. For the sake of simplicity, we will consider only parameter values which result in a single stable limit cycle of period $T=1$ year. 

Similar to linear stability analysis of fixed points, there is a well-developed theory for analysing perturbations around limit cycles. Writing $(S^\ast(t),E^\ast(t),I^\ast(t))$ for the limit cycle, we introduce the vector
\begin{equation}
\bm{\xi}(t)=\sqrt{N}\left(\begin{array}{c}S(t)-S^\ast(t)\\E(t)-E^\ast(t)\\I(t)-I^\ast(t)\end{array}\right)\,,
\end{equation}
where the factor $\sqrt{N}$ has been introduced for notational consistency with later sections. Since we are effectively linearizing the system, it should be noted that this factor has no effect on the following analysis. To first order then, the dynamics of $\bm{\xi}$ are governed by 
\begin{equation}\label{floquetDet}
\frac{d\bm{\xi}}{dt}=J(t)\bm{\xi}\,,
\end{equation}
where $J$ is the time-dependent Jacobian of (\ref{seirDet}). This equation may be solved using Floquet theory \cite{grimshaw}. The key result of the theory states that solution trajectories may be decomposed into the product of a periodic vector with an exponentially growing/decaying amplitude. General solutions are then of the form
\begin{equation}
\bm{\xi}(t)=\sum_{i=1}^nc_i\bm{q}_i(t)e^{\sigma_i t}\,,
\end{equation}
where $n$ is the number of degrees of freedom in the model, $c_i$ is a constant, $\bm{q}_i(t)$ a periodic vector with the same period $T$ as the limit cycle, and the value $\sigma_i$ determining the rate of growth/decay is referred to as the Floquet exponent. Akin to the eigenvalues of the fixed point Jacobian, Floquet exponents are indicative of the stability of the limit cycle in the time-varying directions $\bm{q}_i(t)$; perturbations to the trajectory will grow if $\textrm{Re}[\sigma_i]>0$ and decay if $\textrm{Re}[\sigma_i]<0$.

Although this formalism may be carried through for much of the calculation analytically, ultimately the Floquet exponents and periodic vectors will be obtained numerically. Of course, other techniques exist to study driven systems \cite{coullet,arnoldNormalForm,roberts} and Floquet exponents may be obtained in some cases through asymptotic expansions in small amplitudes, but we wish to align our approach to the previous work on the system which we will investigate below \cite{Ganna}. The procedure (detailed in \ref{floquetBackground}) requires  first computing a matrix whose columns are the independent solutions to  $\eref{floquetDet}$, $X(t)=(\bm{\xi}_1(t) \cdots \bm{\xi}_n(t))$, and then determining the eigenvalues of $X(T)$. If there is a rapid collapse along a stable direction, followed by slow dynamics in a subspace, then the columns of $X(T)$ will be almost linearly dependent, since all solution trajectories quickly move towards the
subspace. In turn, the real parts of eigenvalues of the matrix $X(T)$ will differ by many orders of magnitude. Obtaining these eigenvalues accurately is crucial to the remaining analysis. However, if the disparity between the eigenvalues is too great, numerical procedures may not maintain sufficient accuracy in calculations involving the eigenvalues. This problem was previously highlighted in \cite{Ganna} within an analysis of the seasonally forced SEIR model. There, using the same epidemiological parameters as used in \fref{spirals}, the eigenvalues of $X(T)$ differed by a factor of $10^{59}$. This prompted the authors to implement arbitrary precision numerical methods, at a considerable cost in computing time. We propose that the general method detailed in \Sref{generalFormulation} can be used as a technique to remove this fast direction and hence circumnavigate the numerical difficulties encountered in the analysis.

\subsection{Stochastic treatment exploiting the slow manifold}\label{seirStochastic}

Beginning with the unforced case, we let $\lambda_1,\,\lambda_2$ and $\lambda_3$ be as in (\ref{l1}, \ref{l23}) and write $\bm{v}_1,\,\bm{v}_2,\,\bm{v}_3$ for the corresponding eigenvectors. We introduce the transformation matrix $V=(\bm{v}_1\quad (\bm{v}_2 + \bm{v}_3)\quad i(\bm{v}_2-\bm{v}_3))^T$ and new variables
\begin{equation}
\left(\begin{array}{c}w\\z_2\\z_3\end{array}\right)=V\left(\begin{array}{c}S\\E\\I\end{array}\right)\,.
\label{yzSEI}
\end{equation}
The Jacobian of the transformed system at the endemic fixed point takes the form
\begin{equation}
\bar{J}=\left(\begin{array}{ccc}\lambda_1 & 0 \\ 0 & L\end{array}\right)+\mathcal{O}(\mu^{3/2})\,,\quad\textrm{where}\quad L=\left(\begin{array}{cc} \textrm{Re}[\lambda_2] & \textrm{Im}[\lambda_3]\\ \textrm{Im}[\lambda_2] &\textrm{Re}[\lambda_3]\end{array}\right)\,.
\end{equation}
The nullcline for $w$ is determined by manipulating \eref{nullc} into the form $w=\theta(z_2,z_3)$, with $\bm{A}$ copied from \eref{seirDet}. Although an explicit form for $\theta$ can be found, the expression is too complicated to be worth reproducing here. 

To capture the effects of stochasticity, we introduce variables describing the fluctuations in the new coordinates, rescaled by a factor of $\sqrt{N}$,
\begin{equation}
\bm{\bar{\xi}}=\sqrt{N}\left(\begin{array}{c}w-w^\ast\\\bm{z}-\bm{z}^\ast\end{array}\right)\,.
\end{equation}
Making this substitution in (\ref{seirDet}) and keeping only first order terms in $1/N$ and $\mu$, we find that $\bm{\bar{\xi}}$ obeys
\begin{equation}\label{Jxi}
\frac{d\bm{\bar{\xi}}}{dt}=\bar{J}\bm{\bar{\xi}}+\bm{\zeta}(t)\,,
\end{equation}
where 
\begin{equation}
\left\langle\zeta_i(t)\zeta_j(t')\right\rangle=\delta(t-t')\tilde{B}_{ij}\,, \quad \, \, i,j=1,2,3.
\end{equation}
The matrix $\tilde{B}$ is given by
\begin{equation}
\tilde{B}=V B V^T\bigg|_{(S,E,I)=(S^\ast,E^\ast,I^\ast)}\,,
\end{equation}
where $B$ is as in (\ref{SEIRcorr}). Note that applying the constraint $w=\theta(z_2,z_3)$ induces the relationship
\begin{equation}
w^\ast+\frac{\bar{\xi}_1}{\sqrt{N}}=\theta\left(z_2+\frac{\bar{\xi}_2}{\sqrt{N}},z_3+\frac{\bar{\xi}_3}{\sqrt{N}}\right)\,.
\end{equation}
Expanding once more in large $N$, we find
\begin{equation}
\bar{\xi}_1=\Big[ \bar{\xi}_2\frac{\partial\theta}{\partial z_2}+\bar{\xi}_3\frac{\partial\theta}{\partial z_3} \Big] \Bigg|_{\bm{z}=\bm{z}^\ast}\,.
\label{xi1from23}
\end{equation}
After elimination of the fast direction, \eref{Jxi} becomes
\begin{equation}
\frac{d}{dt}\left(\begin{array}{c}\bar{\xi}_2\\\bar{\xi}_3\end{array}\right)=L\left(\begin{array}{c}\bar{\xi}_2\\\bar{\xi}_3\end{array}\right)+\left(\begin{array}{c}\zeta_2(t)\\\zeta_3(t)\end{array}\right)\,,
\label{xi23sde}
\end{equation}
where $\zeta_2$ and $\zeta_3$ now have correlation matrix $\bar{B}$, which is related to $\tilde{B}$ by \eref{Bpp}.

We move on now to study the situation of seasonally forced infection rate. In principle, the calculations above apply only in the limit of small forcing amplitude (that is, $\delta\to0$ in (\ref{betat})). We learnt in the illustrative ecological example, however, that although our theory is developed to apply in the locality of a stable fixed point, it can continue to provide a useful approximation if this condition does not strictly hold. Applying that lesson to the present case, we modify \eref{xi23sde} to allow $L$ and $\bar{B}$ to become functions of time, as dictated by the replacement $\beta\mapsto\beta(t)$. Essentially, we are approximating the limit cycle by its projection on to the nullcline of the fast direction at the endemic fixed point of the unforced model, and then demanding that any stochastic fluctuations remain on this nullcline. The application of such a coordinate change to the forced system can be further motivated by a consideration of the periodic matrix $J(t)$ in \eref{floquetDet} in the limit of small forcing. While our description of the slow manifold is static, the possibility of a dynamic slow manifold has been explored in the deterministic setting \cite{chicone}.

A Floquet analysis of the deterministic part of the reduced system finds two complex conjugate Floquet multipliers, with the third disparate multiplier having been eliminated from the system. This system no longer suffers from the numerical difficulties which plague the full three-dimensional model. 

To quantify the effect of stochastic fluctuations in this model, we follow the standard procedure of computing the autocorrelation matrix $\mathcal{C}(\tau)$ of oscillations around the limit cycle, which has entries
\begin{equation}
\Big[\mathcal{C}(\tau)\Big]_{ij}= \frac{1}{T}\int^{T}_{0} \Big\langle \bar{\xi}_{i}(t) \bar{\xi}_{j}(t+\tau) \Big\rangle dt\,.
\end{equation}
Of course our reduced system (\ref{xi23sde}) is two-dimensional, meaning that the entries of $\mathcal{C}$ pertaining to $\bar{\xi}_1$ must be deduced from \eref{xi1from23}. The coordinate transformation applied at the start in (\ref{yzSEI}) may then be inverted to give the autocorrelation matrix for the fluctuations in $S$, $E$ and $I$. 

The Fourier transform of the diagonal entries of the autocorrelation gives the power spectrum of oscillations, which provides a convenient visualization of the stochastic fluctuations. In \fref{power} we plot the power spectrum of fluctuations in the number of infected individuals around the limit cycle, comparing the stochastic simulations and the theoretical prediction using the reduced-dimension model (\ref{xi23sde}). The agreement between the simulation and theory can be seen to be excellent; the spectra lie virtually coincident. The peaks in the theoretical spectrum are found at the same positions as those for the simulated spectrum. These are approximately given by 
\begin{equation}\label{psPeaks}
v_{j} = \frac{j}{T} \pm \frac{|\textrm{Im}[\sigma_{1,2}]|}{2 \pi}\, ,
\end{equation}
where $j$ is an integer and $\sigma_{i}$ are the Floquet exponents as defined in \ref{floquetBackground}. This is in agreement with \cite{black,Ganna} where peaks at these positions were also found.
The overall benefit that was garnered by the procedure however was that of computational efficiency. The computation of the theoretical power spectra approximated from the reduced system takes only a small fraction of the computing time of the full system. 

\begin{figure}
\begin{center}
\includegraphics[width=0.8\textwidth, height=0.3\textheight]{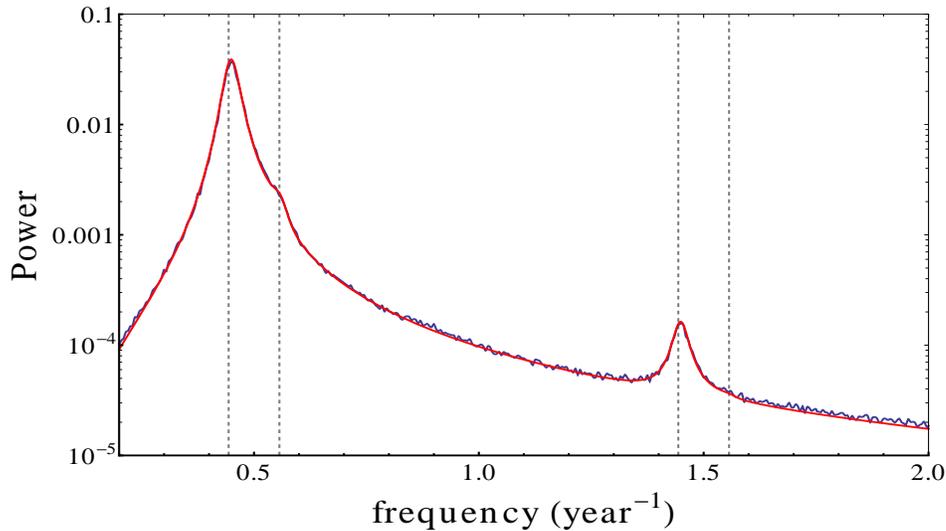}
\end{center}
\caption{Power spectra for the fluctuations in the number of infected individuals about the limit cycle. The analytical power spectrum calculated using the slow manifold approximation is plotted in red, while the power spectrum from stochastic simulations is in blue. Agreement is such that the spectra are difficult to distinguish; the spectrum from simulated results is primarily discernible through its stochastic nature relative to the smooth analytical line. Dotted lines indicate the position of the peaks in the power spectra given by \eref{psPeaks}. Epidemiological parameters are $\beta_{0}=1575$ year$^{-1}$, $\alpha=35.84$ year$^{-1}$, $\gamma=100$ year$^{-1}$ and $\mu=0.02$ year$^{-1}$. The simulated spectrum is calculated as the average power spectrum of $1000$ stochastic realizations, each lasting $200$ years with a system size of $N=10^{8}$.}\label{power}  
\end{figure}

\section{Conclusion}

In this article we have introduced a systematic and general procedure to eliminate fast degrees of freedom in stochastic dynamical systems derived from an individual based model. The method was inspired by the highly successful theory of slow manifolds in deterministic systems, from which we borrow the basic notion of restricting attention to trajectories occupying a low-dimensional subspace. In the stochastic setting, we achieve this by enforcing the condition that the system state remains fixed to the nullcline of the fast direction. We apply this condition to SDE systems with correlated Gaussian white noise, obtaining a lower-dimensional system in which the fast degree of freedom has been eliminated. Importantly, the reduced system has the same basic form as the original, meaning that we have not inadvertently introduced any complicating factors such as non-Markovian processes or coloured noise.

Although our procedure is by no means the only way to reduce the dimension of a stochastic dynamical system, it does offer some distinct advantages. We would argue that our approach is relatively simple: working in the SDE setting makes clear the analogy to deterministic slow manifold theory, and our basic idea (confining the stochastic system to the slow manifold) is intuitively easy to grasp. Moreover, the technique is generally applicable; it does not require an explicitly fast variable, since the fast direction is determined instead from a linear stability analysis. Nor does the technique itself require knowledge of the deterministic trajectories, as is necessary in \cite{Doering}. Finally, as we have shown, the approximation often remains successful even outside of the parameter regime in which it is developed. Why this is so is not understood at present; it appears that the ansatz \eref{ansatz} is surprisingly good at capturing the essential features of the reduced problem. Further work is required to elucidate the limitations of the method, and to characterise the systems to which it can be successfully applied. One restriction is however already apparent: confinement to the slow manifold removes any effects arising from the nature of the flow field away from the manifold. For example, in \cite{Doering} it was shown that the way in which stochastically perturbed trajectories relax back can induce a small flow along the manifold; this effect is overlooked by our method.

Our aim in this paper has been to develop a technique which may be applied to real problems of interest that can be formulated as stochastic dynamical systems. Looking to the future, there is the possibility of developing a general framework which includes the methodology we have described, and we hope that the work presented in this article will provide a significant step towards that goal. 

\ack

We wish to thank G. Rozhnova for useful discussions, and A. J. Roberts for useful correspondence. G.W.A.C. thanks the
Faculty of Engineering and Physical Sciences, University of Manchester for
funding through a Dean's Scholarship. A.J.M and T.R. acknowledge partial 
support through EPSRC grant EP/H02171X/1.  

\appendix

\section{}\label{stochasticBackground}

In this appendix we give the details of the derivation of the SDE systems (\ref{xysde}) and (\ref{SEIsde}), which are mesoscopic descriptions of the individual based models used as examples in sections 2 and 3, respectively. The general formulation of the method is described in \cite{McKaneBiancalaniRogers2013}. 

Recall the model studied in section 2, with reactions listed in \eref{ecomodel}. The state of the system at a given time is specified by the vector $\bm{n}=(n_X,n_Y)$, where $n_X$ and $n_Y$ denote the number of individuals in populations $X$ and $Y$, respectively. We write $P(\bm{n},t)$ for the probability that the model is in state $\bm{n}$ at time $t$. To properly specify the model, we must write a master equation describing the time evolution of $P$. 

The reactions (\ref{ecomodel}) define the possible ways in which the system can transition from the state $\bm{n}'$ to the state $\bm{n}$; the rate with which these transitions occur is written $T(\bm{n}|\bm{n}')$. Specifically, we have 
\begin{eqnarray}
T(n_X+1,n_Y|n_X,n_Y)=(1-\mu)n_X+\mu n_Y\nonumber\\
T(n_X,n_Y+1|n_X,n_Y)=\mu n_X+(1-\mu) n_Y\nonumber\\
T(n_X-1,n_Y|n_X,n_Y)=\varepsilon(1/2-p)n_X^2+\varepsilon(1/2+p) n_Xn_Y\nonumber\\
T(n_X,n_Y-1|n_X,n_Y)=\varepsilon(1/2+p)n_Xn_Y+\varepsilon(1/2-p) n_Y^2\,.
\label{Tnn}
\end{eqnarray}
The master equation is written in terms of the transition rates in the following way;
\begin{equation}
  \frac{\partial P(\bm{n},t)}{\partial t} = \sum_{n} [T(\bm{n}|\bm{n}')(t)P(\bm{n}',t) - T(\bm{n}'|\bm{n})(t)P(\bm{n},t)]\,.
  \label{Meqn}
\end{equation}
While the master equation describes the dynamics of the system entirely, it is in general not analytically tractable. A particular realization of a process obeying the master equation may be simulated using the Gillespie algorithm \cite{gillespie}. While the simulation does not necessarily provide as deep an understanding of the system as an analytical treatment, it provides a useful comparison for the analytical techniques. 

To make analytical progress, we consider the limit of small $\varepsilon$, applying a procedure known as the Kramers-Moyal expansion \cite{vanKampen,McKaneBiancalaniRogers2013}. Briefly, it involves the introduction of the rescaled state vector $\bm{x}=\varepsilon\bm{n}$, followed by a Taylor expansion of $T$ and $P$ around $\bm{x}$. Truncating after two terms, results in a Fokker-Planck equation \cite{Risken} describing the evolution of the probability density function;
\begin{equation}
 \frac{\partial P(\bm{x},t)}{\partial t} = -\sum_{i} \frac{\partial}{\partial x_{i}} \left[ A_{i}(\bm{x})P(\bm{x},t) \right] +\frac{\varepsilon}{2}\sum_{i,j}\frac{\partial^{2}}{\partial x_{i}\partial x_{j}}\left[ B_{ij}(\bm{x})P(\bm{x},t)\right]\,.
\end{equation}
The vector $\bm{A}(\bm{x})$ and the matrix $B(\bm{x})$ can be calculated from the probability transition rates (\ref{Tnn}). It can be shown \cite{Gardiner} that the above equation is equivalent to the SDE
\begin{equation}
 \frac{d\bm{x}}{dt} = \bm{A}(\bm{x}) + \bm{\eta}(t),
\end{equation}
defined in the sense of It\={o} \cite{jacobs}. Here $\bm{\eta}(t)$ is as usual a Gaussian white noise term such that $\langle \bm{\eta}(t) \rangle = 0$ and $\langle \eta_{i}(t)\eta_{j}(t') \rangle = \varepsilon B_{ij}\delta(t-t')$. For the present example, the SDE formulation is given in \eref{xysde}.

The procedure is much the same for the epidemiological model considered in section 4. The transition rates are this time derived from the reactions (\ref{SEIRreactions}), and the state vector is now $\bm{n}=(n_S,n_E,n_I)$. As is usual in this type of model, we choose $1/N$ (where $N$ is the total number of individuals in the population) for the small parameter, $\epsilon$, used in the Kramers-Moyal expansion. Equation \eref{SEIsde} is found as the result. 

In general, we note that the noise matrix $B$ may be decomposed as $B=SRS^T$, where $S$ is a stoichiometric matrix and $R$ a diagonal matrix of rate coefficients \cite{McKaneBiancalaniRogers2013}. Since the rate coefficients are generally positive, we have that $rank(B)=rank(S)$ and thus $B$ is singular if and only if there exists a vector $\bm{v}$ such that $S\bm{v}=0$. This is a linear combination of reactants which is unchanged by any reaction, that is, a conservation relation. Thus the rank of the noise matrix $B$ is typically never less than the number of degrees of freedom in the system.

\section{}\label{singularExample}

We have commented that our method will in general fail to give a reasonable approximation of the dynamics in situations where the noise matrix is singular, but argued that in the case of SDEs developed from an IBM this will almost never be the case (see \ref{stochasticBackground}). However it is perhaps useful to consider a simple example to demonstrate how and why the method fails in the case of a singular noise matrix, in order to highlight its range of validity. We consider then the simple linear SDE system with additive noise,
\begin{eqnarray}
 \frac{d}{dt}\left( \begin{array}{c} x \\  y \end{array} \right) = \left( \begin{array}{cc} 0 & 1 \\ 0 & -1 \end{array} \right) \left( \begin{array}{c} x \\  y \end{array} \right) + \left( \begin{array}{c} \eta_{1}(t) \\   \eta_{2}(t) \end{array} \right),
\end{eqnarray}
with $\eta(t)$ as usual a zero mean Gaussian white noise process delta-correlated in time, with noise correlation matrix,
\begin{eqnarray}
 B= \left( \begin{array}{cc} 0 & 0 \\ 0 & b \end{array} \right),
\end{eqnarray}
which is singular. The noise term $\eta_{1}$ is then identically zero. We proceed through the standard methodology outlined in \sref{generalFormulation} by identifying the eigenvalues and eigenvectors of the system. We find in fact that a centre manifold exists, since
\begin{eqnarray}
\lambda_{1} = -1, \, \, \, \, \, \bm{v}_{1} = \left(\begin{array}{cc}-1\\1\end{array}\right), \, \, \, \, \, \lambda_{2} = 0, \, \, \, \, \, \bm{v}_{2} = \left(\begin{array}{cc}1\\0\end{array}\right).
\end{eqnarray}
The transformed variables are then given by 
\begin{eqnarray}
 \left(\begin{array}{cc}x\\y\end{array}\right) &= \left(\begin{array}{cc} -1 & 1 \\1 & 0  \end{array}\right)\left(\begin{array}{cc}w\\z\end{array}\right), \\
 \left(\begin{array}{cc}w\\z\end{array}\right) &=V\left(\begin{array}{cc}x\\y\end{array}\right),
\end{eqnarray}
where
\begin{eqnarray}
 V &= \left(\begin{array}{cc} 0 & 1 \\1 & 1  \end{array}\right).
\end{eqnarray}
This leads to the dynamical equations
\begin{eqnarray}\label{dynamicsTransformed}
\frac{d}{dt}\left(\begin{array}{cc}w\\z\end{array}\right) = \left(\begin{array}{cc} -1 & 0 \\0 & 0  \end{array}\right)\left(\begin{array}{cc}w\\z\end{array}\right)  + \left(\begin{array}{cc} \tilde{\eta}_{1}(t) \\ \tilde{\eta}_{2}(t) \end{array}\right),
\end{eqnarray}
with noise correlation matrix
\begin{eqnarray}
\tilde{B} = \left(\begin{array}{cc} b & b \\ b &b  \end{array}\right).
\end{eqnarray}
The fast direction is clearly now $w$, with the centre manifold being the plane $w=0$. We therefore condition the noise to lie on this manifold. This yields an equation for $z$
\begin{equation}
 \frac{dz}{dt}= 0 
\end{equation}
since the conditioned noise correlator is
\begin{equation}
 \bar{B} = \tilde{B}_{22} - \frac{\tilde{B}_{12} \tilde{B}_{21} }{ \tilde{B}_{11} } = 0.
\end{equation}
Our method therefore suggests that there are no dynamics, deterministic or stochastic, in $x=z-w$. Following this example it is clear why the method fails; the noise is conditioned to lie on the nullcline. If the noise correlator $B$ is singular, conditioning the noise to one dimension effectively kills off any stochastic dynamics. It is perhaps interesting to note that this was also the result of the averaging approximation, as highlighted in \cite{roberts}.

\section{}\label{floquetBackground}

The analogue of a linear stability analysis for systems with periodic components is known as Floquet theory \cite{grimshaw}. It can also play an important role in the analysis of stochastic fluctuations about a deterministic trajectory \cite{black,Ganna}. In this appendix the general formulation of Floquet theory is discussed before the more detailed application to linear stochastic systems is given.

Floquet theory gives the solutions to sets of linear differential equations in the form of \eref{floquetDet}, where $J(t)$ is periodic with a period $T$. The general solution can be shown to be 
\begin{equation}
\bm{\xi}(t)=\sum_{i=1}^nc_i\bm{q}_i(t)e^{\sigma_i t},
\end{equation}
where $\bm{q}_{i}(t)$ is a periodic vector and $\sigma_{i}$ are termed the Floquet exponents of the system. The quantities $\rho_{i}=e^{\sigma_{i}T}$ are termed the Floquet multipliers of the system.

In particular one can work in a canonical form for calculational ease, with canonical quantities denoted with a superscript $0$. The canonical form is constructed from $n$ decomposed solutions to \eref{floquetDet} such that $\bm{\xi}^{(0)}_{i}(t) = \bm{q}^{(0)}_i(t)e^{\sigma_i t} $. A fundamental matrix of these solutions may then be introduced along with matrices $Y^{(0)}$ and $Q^{(0)}$. For the case $n=3$ these may be expressed as
\begin{eqnarray}
 X^{(0)}=[\bm{\xi}^{(0)}_{1}(t),\bm{\xi}^{(0)}_{2}(t),\bm{\xi}^{(0)}_{3}(t)],\\
 X^{(0)}=Q^{(0)}Y^{(0)},\\
 Q^{(0)}=[\bm{q}^{(0)}_{1}(t),\bm{q}^{(0)}_{2}(t),\bm{q}^{(0)}_{3}(t)],\\
 Y^{(0)}=Diag[e^{\mu_{i}t}].
\end{eqnarray}
A method for obtaining the Floquet multipliers $\mu_{i}$ along with the canonical form of the solutions is now required. Obtaining both is dependent on the determination of a matrix known as the  monodromy matrix, which we shall now discuss. 

The monodromy matrix, $D$, is defined such that $X(t+T)=X(t)D$, for any fundamental matrix $X(t)$ constructed from linearly independent solutions to \eref{floquetDet}. It can be shown that while the monodromy matrix is dependent on the fundamental matrix chosen, its eigenvalues are not \cite{grimshaw}. The eigenvalues of $D$ are $\rho_{i}$, the Floquet multipliers of the system. Further, if a matrix $W$ is constructed from the eigenvectors of $D$, the canonical fundamental matrix $X^{(0)}(t)$ is related to a general fundamental matrix $X(t)$ via $X^{(0)}(t) = X(t) W$. Therefore, the monodromy matrix allows the canonical fundamental matrix $X^{(0)}(t)$ to be determined from a general fundamental matrix $X(t)$, along with the matrix $Y^{(0)}$. From these the periodic matrix $Q^{(0)}(t)$ can also be deduced. 

In general, once the fundamental matrix is obtained it will have to be transformed into canonical form by a numerical determination of the monodromy matrix, $D=X^{-1}(t)X(t+T)$. For a system with initial conditions $t=0$, $X(0)=I$, this simplifies to $D = X(T)$.

Now the stochastic system can be considered;
\begin{equation}
\frac{d\bm{\xi}}{dt}=J(t)\bm{\xi} + \bm{\eta}(t)\, ,
\end{equation}
where $\bm{\eta}(t)$ is a vector of Gaussian white noise terms defined as in \ref{stochasticBackground}, except that now the noise covariance matrix depends explicitly on time through the varying parameter $\beta(t)$; $\langle \eta_{i}(t)\eta_{j}(t') \rangle = \varepsilon B_{ij}(t)\delta(t-t')$. The solution may be constructed as a sum of the general solution to \eref{floquetDet} along with a particular solution, so that
\begin{equation}
 \bm{\xi}(t) = X^{(0)}(t)\bm{\xi}^{(0)} + X^{(0)}(t) \int\limits_{t_{0}}^{t} \left[X^{(0)}(s)\right]^{-1}\bm{\eta}(s)ds,
\label{floqPSStep1}
\end{equation}
or, setting the initial conditions in the infinite past and making a change of integration variable $s \rightarrow s'=t-s$
\begin{equation}
 \bm{\xi}(t) = Q^{(0)}(t) \int\limits_{t_{0}}^{t} Y^{(0)}(s')\left[Q^{(0)}(t-s')\right]^{-1}\bm{\eta}(t-s')ds'.
\label{floqPSStep2}
\end{equation}
In the course of the analysis conducted in \Sref{seirSection}, $\bm{\xi}(t)$ represents some stochastic fluctuation around limit cycle behaviour. An obvious quantity of relevance is the power spectrum of such fluctuations. To obtain the power spectrum, one first calculates the two-time correlation function $C(t+\tau,t) = \langle \bm{\xi}(t+\tau) \bm{\xi}^{T}(t) \rangle $; substituting \eref{floqPSStep2} one obtains
\begin{equation}
 C_{ij}(t+\tau,t)=Q^{(0)}(t+\tau)Y^{(0)}(\tau)\Lambda(t)\left[Q^{(0)}(t)\right]^{T}, \label{floqPSStep4}
\end{equation}
with 
\begin{equation}
\Lambda(t)= \int\limits_{t_{0}}^{\infty}Y^{(0)}(s)\Gamma(t-s)Y^{(0)}(s)ds \label{floqPSStep5}
\end{equation}
and 
\begin{equation}
\Gamma(s) =\left[ Q^{(0)}(s) \right]^{-1} B(s) \left[ \left[Q^{(0)}(s)\right]^{-1} \right]^{T}. \label{floqPSStep6}
\end{equation}
The correlation function, $\mathcal{C}(\tau)$ is then simply related to the two-time correlation function by
\begin{equation}
 \mathcal{C}(\tau) = \frac{1}{T}\int\limits_{0}^{T} C(t+\tau,t) dt.
\end{equation}
In turn, the Wiener-Khinchin theorem tells us that the power spectrum, $P(\omega)$, is simply the Fourier transform of the correlation function, and so
\begin{equation}
P_{i}(\omega)=\int{\mathcal{C}_{ii}(\tau)e^{i\omega\tau}d\tau}.\\ \label{floqPSStep3}
\end{equation}
The intermediate steps are left to the reader, but full details are found in \cite{boland}. A key point to note is that Equations (\ref{floqPSStep1}-\ref{floqPSStep6}) hold \textit{only} for the canonical matrices $X^{(0)}$, $Q^{(0)}$ and $Y^{(0)}$. 

\section*{References}


\providecommand{\newblock}{}

\end{document}